\documentclass{aa}
\usepackage{times,epsf,psfig}
\begin{document}

   \thesaurus{09
             (09.10.1;
               08.06.2;
               09.09.1 HH 46/47)}
   \title{HH\,46/47: Also a parsec scale flow 
	\thanks{Based on observations collected at the European Southern Observatory, Chile (ESO No.\ 62.I-0848)}}

   \author{Thomas Stanke
	\and Mark J. McCaughrean
	\and Hans Zinnecker}

   \offprints{Th.\ Stanke}

   \institute{Astrophysikalisches Institut Potsdam (AIP),
              An der Sternwarte 16, D-14482 Potsdam\\
              email: tstanke@aip.de
%         \and
%             University of Alexandria, Department of Geography\\
%             email: c.ptolemy@hipparch.uheaven.space
%             \thanks{The university of heaven temporarily does not
%                     accept e-mails}
             }

   \date{Received; accepted}

   \maketitle

\sloppy

   \begin{abstract}
We report the discovery of a pair of large Herbig-Haro type structures roughly
10 arcminutes (1.3\,pc) north-east and south-west of the source driving 
the well-known \object{HH\,46/47} Herbig-Haro jet in new deep emission-line images
made using the Wide Field Imager on the ESO/MPG La Silla 2.2-m telescope.
These new images suggest that the \object{HH\,46/47} outflow is much more extensive
than previously assumed, extending over a total of 2.6\,pc on the sky, or
over 3\,pc in space, when deprojected. \object{HH\,46/47} thus also belongs to the 
recently-discovered class of giant Herbig-Haro flows. 
 
   \end{abstract}

%
%________________________________________________________________

\section{Introduction}
Observations over the past few years with large sensitive imaging cameras
have shown that outflows and jets from young stars often extend over several 
parsecs, much further than previously suspected. Examples of these
giant flows include well-known jets such as \object{HH\,34}, \object{HH\,24}, 
\object{HH\,111}, \object{HH\,1/2}, and the flow from \object{T\,Tau} 
(e.g.\ Bally \& Devine 1994, 1997; Devine et al.\ 
1997; Eisl\"offel \& Mundt 1997; Reipurth et al.\ 1997, 1998b). The discovery 
that such outflows can be so large has significant implications for our
understanding of their lifetimes and the cumulative impact they may have 
on their natal molecular clouds.

\object{HH\,46/47} is a well-studied prototypical Herbig-Haro jet located in the 
\object{Gum Nebula}, driven by a young source (\object{HH\,47\,IRS} or \object{IRAS\,08242$-$5050}; 
Graham \& Elias 1983; Emerson et al.\ 1984; Cohen et al.\ 1984; Graham \& 
Heyer 1989, Reipurth \& Heathcote 1991) embedded in a southern Bok globule 
(\object{GDC\,1} =~\object{ESO\,210$-$6A}; Bok 1978; Reipurth 1983). The two brightest 
components of this outflow system, \object{HH\,46} and \object{HH\,47}, were discovered by 
Schwartz (1977), and shown by Dopita et al.\ (1982; see also Graham \& 
Elias 1983) to be shock-excited HH objects in a collimated bipolar flow.
The blue-shifted lobe of the flow, with main components \object{HH\,46}, 
\object{HH\,47\,A}, and \object{HH\,47\,D}, is prominent in optical 
emission as it escapes the globule, whereas 
the red-shifted lobe is mostly hidden from view as it runs through the 
globule, only to become visible again in the guise of \object{HH\,47\,C}
as it leaves the globule; H$_2$ emission from the jet as it traverses the 
globule can however be seen in the near-infrared (Eisl\"offel et al.\ 1994). 
From \object{HH\,47\,D} in the north-east to \object{HH\,47\,C} in the 
south-west, the flow was thought 
to extend over 0.57\,pc in projection at a distance of $\sim$\,450\,pc.  

Ground-based optical imaging has been presented by Bok (1978), Reipurth \& 
Heathcote (1991), Hartigan et al.\ 1993, Eisl\"offel \& Mundt (1994), and 
Morse et al.\ (1994) among others, while HST images have been presented by 
Heathcote et al.\ (1996). A characteristic feature of the \object{HH\,46/47}
jet is 
the apparent change in flow direction with time, from small-scale wiggles 
seen in the HST images (Heathcote et al.\ 1996) to a more gradual turning 
towards smaller position angles (Reipurth \& Heathcote 1991). There is
also an obvious misalignment of $\sim$17$\degr$ between the jet and the 
counterjet seen near the driving source (Reipurth \& Heathcote 1991).  

Morse et al.\ (1994) found that the outermost bowshock, \object{HH\,47\,D},
is running into previously accelerated material and they presented an image 
showing an H$\alpha$ feature 50 arcsec ahead of \object{HH\,47\,D}, 
which they suggested showed that \object{HH\,46/47} extends over a greater 
length than previously known. In this paper, we report the discovery of two 
groups of Herbig-Haro type objects at distances of approximately 
10\,arcmin (1.3\,pc) to the north-east and south-west of \object{HH\,47\,IRS} 
and as a result, we argue that the \object{HH\,46/47} 
flow is indeed much larger than previously suspected, extending over several 
parsecs at least.

\begin{figure*}
	\vspace{0cm}
%\centerline{\hspace{0cm}\epsfxsize=18cm \epsfbox{hh46fig1.eps}}
	\vspace{0cm}
	\caption{Narrow-band 658\,nm (H$\alpha$+continuum) image of the area 
        around \object{GDC\,1}, the globule harbouring the \object{HH\,46/47}
	energy source, \object{HH\,47\,IRS} (position marked by the cross). 
	The image covers 21\arcmin\,$\times$14\arcmin, i.e.\ 
	$\sim$2.7\,pc\,$\times$1.8\,pc at a distance of 450\,pc. The globules
	\object{GDC\,1} and~2 are seen as opaque, dark 
        features with bright rims at their north-western edges. Several 
        smaller, less opaque globules are outlined by bright rims, including 
        a very narrow, transparent fingerlike structure. In addition to the 
        overall diffuse H$\alpha$ emission from the \object{Gum Nebula}, 
	compact structures are seen, namely the known \object{HH\,46/47}
	jet complex originating in \object{GDC\,1} and the newly discovered 
	HH objects (marked by arrows).
	}
	\label{hh46ha}
\end{figure*}

\begin{figure*}
	\vspace{0cm}
%\centerline{\hspace{0cm}\epsfxsize=18cm \epsfbox{hh46fig2.eps}}
	\vspace{0cm}
	\caption{Continuum subtracted H$\alpha$ image of the same area as in 
        Fig.\ \ref{hh46ha}. The major features in the \object{HH~46/47} jet 
	are labeled. The feature ahead of \object{HH\,47\,D} found by 
	Morse et al.\ (1994) is also seen. The newly discovered HH objects 
	are labeled as \object{HH\,47\,NE} and \object{HH\,47\,SW}\@. 
	While \object{HH\,47\,NE} seems to form a large bow shock, the 
        morphology of \object{HH\,47\,SW} is not well defined.}
	\label{hh46hausm}
\end{figure*}

%__________________________________________________________________

\section{Observations and data reduction}
The images presented here were taken with the Wide Field Imager on the 
MPG/ESO 2.2\,m telescope on La Silla on Jan.\ 21 1999. The camera uses
a mosaic of 8 CCDs, each a 2k$\times$4k pixel array, providing a 
total detector size of 8k$\times$8k pixels with a field-of-view of 
$\sim$0.54$\times$0.54\,degrees at an image scale of 0.25 
arcsec/pixel. 
Images were taken through narrow-band filters at 658\,nm (H$\alpha$) and 
676\,nm ([\ion{S}{ii}]), and a medium-passband filter at 753\,nm, which 
was used to measure continuum emission. Integration times were 20\,min 
(H$\alpha$), 30\,min ([\ion{S}{ii}]), and 12.5\,min (continuum), each split 
into 5 dithered exposures to compensate for the gaps between the CCDs, bad 
pixels, and cosmics.

Data reduction was standard, starting with bias frame subtraction and division 
by a flat field frame (a combination of dome and sky flats); bad pixels were 
masked and excluded from further processing. Then mosaics were made from the 
single exposures as follows. First, a ``positional reference frame'' was
constructed from the continuum exposures by accurately registering and 
combining the data for each chip separately, resulting in one average image 
for each chip. Large enough dithering steps had been chosen such that these 
images overlapped, allowing an accurate determination of the position of each 
chip with respect to the others. The averaged images for each of the 8 chips
were then registered, rotated, shifted, and finally averaged into one large 
master image which served as positional reference frame for registering the 
other exposures. All individual exposures through all three filters were then 
registered to this positional reference frame. The final mosaics were then 
constructed by taking the median of the rotated and shifted single exposures 
to reject cosmic ray events.

Further processing of the H$\alpha$ image was applied to show the 
newly discovered HH objects more clearly. The H$\alpha$ frame was
lightly gaussian smoothed so that the PSF matched that of the continuum frame;
the latter was then appropriately scaled and subtracted from the H$\alpha$
image. Since the wavelength difference between two filters was quite large, 
different colours of stars led to many small residual images which were
removed by hand. Finally, a large-scale NW-SE gradient in the H$\alpha$
emission was subtracted from the image to enhance the contrast. 

% The resulting
% image is shown in Fig.\ \ref{hh46hausm}.

%__________________________________________________________________

%__________________________________________________________________

\section{Results}
Figure~\ref{hh46ha} shows a 21\arcmin\,$\times$\,14\arcmin{} 
($\sim$2.7\,pc\,$\times$\,1.8\,pc) section of the H$\alpha$ image. 
The position of the young star driving the \object{HH\,46/47} flow (the only
protostellar object known in the area) is marked
by a cross. It is seen to be embedded in a globule (\object{GDC\,1}) which 
is illuminated and evaporated from the north-west by the exciting stars
of the Gum nebula \ion{H}{ii} region 
\object{$\zeta$\,Pup} and \object{$\gamma^2$\,Vel} several degrees (20-40\,pc)
to the north-west.
In the south-eastern part of the image,
a number of apparently connected illuminated globules, rims, and fingerlike 
structures can be seen: these may be the remains of a larger cloud that
has been evaporated from the north-west. The edges of these globules are
outlined by rims of H$\alpha$ emission. Smooth extended H$\alpha$ emission
is seen to fill the spaces between the globules and to the north-west. The 
\object{HH\,46/47} jet system consists of much more compact, knotty, and 
filamentary H$\alpha$ structures. Superimposed on the smooth background, 
a number of additional compact, knotty patches of H$\alpha$ emission 
can also be seen in the north-eastern and south-western corners of the 
image (marked with arrows). 

\topsep0mm

Figure~\ref{hh46hausm} shows the same field with the continuum emission 
and large-scale H$\alpha$ gradient subtracted in order to show the compact 
H$\alpha$ features better. The main features are the rims around 
the globules whose structure suggests illumination from the north-west, with 
bright emission at their north-eastern tips and streamers 
extending to the south-east. In addition to \object{GDC\,1} harbouring 
\object{HH\,47\,IRS} and a comparably-sized second globule, 
\object{GDC\,2}, at the eastern edge of the 
image (Reipurth 1983), many smaller and less pronounced features are seen, 
including a very narrow finger pointing to the north-west.

The features in the north-east and south-west corners marked by arrows
in Fig.\ \ref{hh46ha} also stand out clearly now. We argue that these are 
Herbig-Haro type structures as follows:
\begin{itemize}
\itemsep0mm
\item The knots are emission line features, detected in H$\alpha$ and in 
      [\ion{S}{ii}] (albeit only faintly in the latter line),  typical 
      of Herbig-Haro objects;
\item They display a much more compact structure than the emission from the
      background \ion{H}{ii} region, suggesting emission from shocked gas
      rather than from the diffuse ionized gas;
\item Their morphology differs from that of the rims around the globules: 
      the latter are indicative of thin layers of gas evaporating from the
      denser globules, wrapping around the globules, while the candidate 
      HH objects show up as compact blobs unassociated with any globules;
      the north-eastern group forms a bow-like structure heading north-east,
      whereas the rims around the globules head north-west;
\item The morphology of the north-eastern group is reminiscent of a large
      fragmented bow shock, indicating a working surface in a flow heading 
      north-east, pointing back to somewhere in the vicinity of 
      \object{HH\,47\,IRS}; 
\item The south-western group is at the same distance from 
      \object{HH\,47\,IRS} as
      the north-eastern group, suggesting that these may be matching shocks
      in a single flow, created by the same ejection event. 
\end{itemize}

We will henceforth refer to the north-eastern and south-western groups of 
knots as \object{HH\,47\,NE} and \object{HH\,47\,SW} respectively. The 
approximate location of \object{HH\,47\,NE} is 8$^{\it h}$ 26$^{\it m}$ 
40$^{\it s}$, $-$50$\degr$ 58$^\prime$
15$^{\prime\prime}$ (J2000), some 9.9\,arcmin ($\sim$1.3\,pc) from 
\object{HH\,47\,IRS} at a position angle of $\sim$74$\degr$. 
\object{HH\,47\,SW} is at roughly 8$^{\it h}$ 24$^{\it m}$ 55$^{\it s}$, 
$-$51$\degr$ 07$^\prime$ 00$^{\prime\prime}$ (J2000),
10.5 arcmin from \object{HH\,47\,IRS} at a position angle of $\sim$50$\degr$.

The discovery of these HH~objects suggests that the previously known 
\object{HH\,46/47} jet is only the innermost portion of a much larger 
flow, as suggested by Morse et al.\ (1994) based on their analysis of the 
\object{HH\,47\,D} bow shock. The flow is now seen to extend over at least 
$\sim$2.6\,pc in projection, or $\sim$3\,pc when deprojected using 
an orientation of $\sim$30$\degr$ out of the plane of the sky (see Eisl\"offel
\& Mundt 1994; however, we do not know if and how the flow direction changes
with respect to the plane of the sky). Thus \object{HH\,46/47} can be added
to the growing class of parsec-scale outflows from young low-mass stars. 
Assuming a tangential flow velocity of about 150 km/s (see, e.g.\ Reipurth 
\& Heathcote 1991; Schwartz et al.\ 1984; Eisl\"offel \& Mundt 1994; Micono 
et al.\ 1998), the dynamical timescale for the new outer bow shocks 
\object{HH\,47\,NE} and SW is about 9000 years. 

\object{HH\,47\,NE} is displaced from the current jet axis (as defined by the 
north-eastern lobe of \object{HH\,46/47} with respect to \object{HH\,47\,IRS},
at a position angle of $\sim$54$\degr$) by about 20$\degr$. However, 
Reipurth \& Heathcote (1991) found that, in addition to short timescale 
kinks and wiggles, the north-eastern lobe of the \object{HH\,46/47} 
jet appears to have been gradually changing its flow direction towards 
smaller position angles, by about 3$\degr$ from \object{HH\,47\,D} to 
\object{HH\,47\,B}\@. The time between the ejection of 
these features is estimated to be about 1300 years; extrapolating this over 
the 9000 year dynamical timescale of \object{HH\,47\,NE} yields a change 
in flow direction of about 21$\degr$. Thus, the observed displacement of 
\object{HH\,47\,NE}
from the current flow axis is consistent with a steady change of flow 
direction over the last 9000 years at the rate currently observed for the 
inner part of the jet. 

While \object{HH\,47\,SW} is only a few degrees off the current axis of 
the flow as defined by the north-eastern lobe of the inner \object{HH\,46/47}
jet, it lies some 8$\degr$ off the axis of the south-western lobe 
(PA$\sim$58$\degr$), as defined by \object{HH\,47\,C} and the counterjet seen 
near the source in [\ion{S}{ii}] and H$_2$. This also indicates a 
gradual change in flow direction in the south-western lobe, albeit smaller 
than seen in the north-eastern lobe.

The different position angles of \object{HH\,47\,NE} and SW also indicate 
that the flow directions of the north-eastern and south-western lobe were 
not coincident at the time \object{HH\,47\,NE} and SW were ejected. This is 
consistent with the observation that the previously known inner jet and 
counterjet are not aligned at their origin (Reipurth \& Heathcote 1991).
The newly discovered binary nature of the outflow source (Reipurth, pers.\ comm.) may hold
the key to understanding the wiggles, bends, and misalignment of the
\object{HH\,46/47} jet and counterjet.
  
\object{HH\,47\,NE} and SW are rather faint in the [\ion{S}{ii}] filter, 
making it difficult to constrain their excitation mechanism. Their 
surface brightness seems to be of the same order as that of the leading bow 
in \object{HH\,47\,D}, while \object{HH\,47\,D} is brighter in H$\alpha$ 
than \object{HH\,47\,NE} and SW\@. This implies that the 
[\ion{S}{ii}]:H$\alpha$ line ratio is greater for
\object{HH\,47\,NE} and SW than for \object{HH\,47\,D}, suggesting that like 
\object{HH\,47\,D}, \object{HH\,47\,NE} and SW are excited by shocks rather
than by external ionization, consistent with the bow shock appearance of 
\object{HH\,47\,NE}\@. However, the signal-to-noise is rather poor in both
images and there are uncertainties in the transmission curve of the 
[\ion{S}{ii}] filter used in the Wide Field Imager. Thus we cannot yet exclude 
the possibility that part of the excitation is due to external irradiation, 
as recently found for some jets in Orion (Reipurth et al.\ 1998a). This 
would not be an unreasonable finding, since \object{HH\,47\,NE} and SW are 
located in the \object{Gum Nebula} \ion{H}{ii} region and thus very likely 
exposed to the ionizing radiation of stars including \object{$\zeta$\,Pup}
and \object{$\gamma^2$\,Vel} responsible for exciting the \ion{H}{ii} region.
Clearly, spectroscopic observations are required to address the excitation 
mechanism of \object{HH\,47\,NE} and SW directly.

In summary, we have found two groups of Herbig-Haro type features, 
\object{HH\,47\,NE} and \object{HH\,47\,SW}, which appear to delineate the 
flow driven by \object{HH\,47\,IRS} over a significantly greater length 
than previously suspected ($\sim$3\,pc), increasing the dynamical age of the 
system to $\sim$\,9000 years.
The positions of \object{HH\,47\,NE} and SW with respect to the driving source
and the inner jet confirm that there have been long-term secular changes 
in flow direction.

\begin{acknowledgements}
      Thanks are due to the ESO staff (in particular Thomas Augusteijn, 
      Dietrich Baade, Pablo Prado, Felipe Sanchez) for their support
      during the observations.
      This work was supported under
      \emph{Deut\-sche For\-schungs\-ge\-mein\-schaft\/} project
      number Zi 242/9-1.
\end{acknowledgements}

%\listofobjects

\begin{thebibliography}{}

\bibitem[1994]{ballydev94} Bally J., Devine D., 1994, ApJ 428, L65

\bibitem[1997]{ballydev97} Bally J., Devine D., 1997, in: Herbig--Haro Flows and the Birth of Low-Mass Stars, IAU Symposium No.\ 182, eds.\ B.\ Reipurth, C. Bertout, Kluwer Academic Publishers, p.\ 29 

\bibitem[1978]{bok78} Bok B., 1978, PASP 90, 489

\bibitem[1984]{cohenetal84} Cohen M., Schwartz R.D., Harvey P.M., Wilking B.A., 1984, ApJ 281, 250

\bibitem[1997]{devineetal97} Devine D., Bally J., Reipurth B., Heathcote S., 1997, AJ 114, 2095

\bibitem[1982]{dopitaetal82} Dopita M.A., Schwartz R.D., Evans I., 1982, ApJ 263, L73

\bibitem[1994]{eisletal94} Eisl\"offel J., Davis C.J., Ray T.P., Mundt R., 1994, ApJ 422, L91

\bibitem[1994]{eislmu94} Eisl\"offel J., Mundt R., 1994, A\&A 284, 530

\bibitem[1997]{eismundt97} Eisl\"offel J., Mundt R., 1997, AJ 114, 280

\bibitem[1984]{emerson84} Emerson J.P., Harris S., Jennings R.E., Beichman C.A., Baud B., Beintema D.A., Marsden P.L., Wesselius P.R., 1984, ApJ 278, L49

\bibitem[1983]{grahamelias83} Graham J.A., Elias J.H., 1983, ApJ 272, 615

\bibitem[1989]{grahamheyer89} Graham J.A., Heyer M.H., 1989, PASP 101, 573

\bibitem[1993]{hartiganetal93} Hartigan P., Morse J.A., Heathcote S., Cecil G., 1993, ApJ 414, L121

\bibitem[1996]{heath96} Heathcote S., Morse J.A., Hartigan P., Reipurth B., Schwartz R.D., Bally J., Stone J.M., 1996, AJ 112, 1141

\bibitem[1998]{miconoetal98} Micono M., Davis C.J., Ray T.P., Eisl\"offel J., Shetrone M.D., 1998, ApJ 494, L227

\bibitem[1994]{morseetal94} Morse J.A., Hartigan P., Heathcote S., Raymond J.C., Cecil G., 1994, ApJ 425, 738

\bibitem[1983]{reipurth83} Reipurth B., 1983, A\&A 117, 183

\bibitem[1997]{reipetal97} Reipurth B., Bally J., Devine D., 1997, AJ 114, 2708

\bibitem[1998]{reipetal98a} Reipurth B., Bally J., Fesen R.A., Devine D., 1998a, Nature 396, 343

\bibitem[1998]{reipetal98b} Reipurth B., Devine D., Bally J., 1998b, AJ 116, 1396 % Herbig Haro flows from the L1641-N embedded infrared cluster

\bibitem[1991]{reipheath91} Reipurth B., Heathcote S., 1991, A\&A 246, 511

\bibitem[1977]{schwartz77} Schwartz R.D., 1977, ApJ 212, L25

\bibitem[1984]{schwartzetal84} Schwartz R.D., Jones B.F., Sirk M., 1984, AJ 89, 1735

%\bibitem[1999]{stanke99} Stanke T., McCaughrean M.J., Zinnecker H., 1999, in prep.\

\end{thebibliography}
\end{document}